%% file: main.tex
\documentclass[conference]{IEEEtran}
\IEEEoverridecommandlockouts
\usepackage{subcaption} 
\usepackage{cite}
\usepackage{amsmath,amssymb,amsfonts}
\usepackage{algorithmic}
\usepackage{graphicx}
\usepackage{textcomp}
\usepackage{xcolor}
\usepackage{tikz}
\usepackage{color}
\usepackage{caption}
\usetikzlibrary{shapes, positioning} 
\usepackage[hidelinks]{hyperref}

\usepackage[dvipsnames]{xcolor}

\def\BibTeX{{\rm B\kern-.05em{\sc i\kern-.025em b}\kern-.08em
    T\kern-.1667em\lower.7ex\hbox{E}\kern-.125emX}}

\def\BibTeX{{\rm B\kern-.05em{\sc i\kern-.025em b}\kern-.08em
    T\kern-.1667em\lower.7ex\hbox{E}\kern-.125emX}}

\definecolor{stripA}{HTML}{000000}
\definecolor{stripB}{HTML}{f2f2f2}

\usepackage{fancyhdr}

\fancypagestyle{firstpage}{
  \fancyhf{}

\fancyfoot[C]{%
  \footnotesize
    Accepted at SC 2025's Women in HPC (WHPC) workshop. The definitive version is available under DOI \href{https://doi.org/10.5281/zenodo.17550110}{10.5281/zenodo.17550110}.
  }

}

\begin{document}

\title{ Mixed-Precision For Energy Efficient Computations%
}



\author{
\IEEEauthorblockN{Gülçin Gedik\IEEEauthorrefmark{1}\IEEEauthorrefmark{2}\IEEEauthorrefmark{3}}
\IEEEauthorblockA{ 
\IEEEauthorrefmark{2}Université Paris-Saclay, UVSQ\\ 
Dresden, Germany \\
gulcin.gedik@mailbox.tu-dresden.de\\
}
\and
\IEEEauthorblockN{Robert Schöne\IEEEauthorrefmark{3}}
\IEEEauthorblockA{
\IEEEauthorrefmark{3}ZIH, CIDS, TU Dresden \\
Dresden, Germany \\
robert.schoene@tu-dresden.de\\
}
\and
\IEEEauthorblockN{Roman Iakymchuk\IEEEauthorrefmark{1}}
\IEEEauthorblockA{
\IEEEauthorrefmark{1}Ume\aa{} University\\
Ume\aa{}, Sweden \\
roman.iakymchuk@umu.se 
}
}

\maketitle
\thispagestyle{firstpage}

\begin{abstract}
As simulations grow more realistic, the pursuit of higher accuracy results in extended computation times and substantial power consumption. 
This study explores mixed-precision computing as a promising strategy to address these challenges, leveraging computer arithmetic tools to optimize performance.

Using Reactor Simulator and LULESH benchmarks as case studies, we evaluated the potential of mixed-precision strategies to reduce both time-to-solution and energy-to-solution. For Reactor Simulator, we achieved a 30\% reduction in both metrics without compromising accuracy. Similarly, for LULESH, results demonstrated up to a 30\% improvement in time-to-solution and a 25\% reduction in energy-to-solution. 
\end{abstract}


\input{intro}


\input{methodology}

\input{reactor}
\input{lulesh}
\input{conclusion}

\clearpage

\section*{ACKNOWLEDGMENTS}

The author gratefully acknowledges the financial support provided by -------- under Grant No. -----, funded by ------. The author would like to thank ------ for supporting this work within -----. The author would like to thank ----- for their valuable guidance and contributions.


\bibliographystyle{ieeetr}
\bibliography{biblio}

\appendices
\section{Measurement Methodology Details}
\input{measurement-meth-details}
\section{Error Analysis Details}
\input{error-analysis-details}

\end{document}

%% file: intro.tex
\vspace*{-3mm}

\section{Introduction and Background}
Recent efforts focused on designing applications that not only deliver high accuracy but also minimize both time-to-solution and energy-to-solution \cite{malms_strategic_agenda,  ExascaleComputingandDataHandlingChallengesandOpportunitiesforWeatherandClimatePrediction,deepseekai2025deepseekv3technicalreport-etal}.



Mixed-precision computing offers potential gains in both time-to-solution and energy-to-solution. However, achieving those gains without compromising scientific accuracy is nontrivial \cite{CHEN2026107990, kashi2025mixedprecisionnumericsscientificapplications}, thus posing a multi-objective optimization challenge.
Our work addresses this challenge by exploring the optimization space defined by hardware capabilities, algorithmic choices, and the selective application of mixed-precision techniques. 

\input{verificarlo-explain}
Exascale computing demands an understanding of how hardware limits, algorithm design, and precision choices interact to impact performance. When these factors are balanced well, applications can be both energy-efficient and reliable. We present our methodology in the following sections and apply it to two case studies, both of which employ explicit solvers and exhibit distinct computational characteristics.



%% file: verificarlo-explain.tex
In parallel with the theoretical advances in floating point arithmetic error mitigation and estimation techniques\cite{highamproba, arar2023}, practical tools for exploring precision and error propagation have emerged. One such tool is \textit{Verificarlo}, which operates at the compiler’s intermediate representation level to instrument and analyze floating-point operations without requiring modifications to the source code. Verificarlo leverages two complementary backends: the MCA backend \cite{verificarlo_mca} to assess error sensitivity in targeted code regions by evaluating numerical stability under randomized perturbations, and the VPREC backend \cite{verificarlo_vprec} to explore mixed-precision strategies, quantifying rounding-error effects and determining the minimal precision required to preserve accuracy or convergence.  

%% file: methodology.tex




\section{Methodology }
\label{sec:methodology}

Changing an application from double to mixed (including lower) precision requires balancing accuracy and performance. Since each application is unique, our approach has to be adaptable. A key challenge is identifying routines that can safely use reduced precision without compromising accuracy or degrading performance \cite{deoliveiracastro2022}. This involves weighing the performance gains of lower precision against the overhead of copying and casting between variable types. Achieving this balance demands a deep understanding of the application's architecture, including data structures, computations, libraries, and communication patterns. 
To guide this, we follow the methodology proposed in \cite{chen2024}. 

We begin by profiling performance hotspots to identify regions with high potential speedup. Then, we analyze numerical hotspots by tracking variables and quantifying error growth with the help of Verificarlo's VPREC and MCA backends. 
We then implement mixed precision to the most time-consuming regions that contribute minimal error, maximizing speedup with minimal accuracy loss. 
This function-level analysis illustrates the workflow and performance-energy trade-offs on sequential versions and also directly extends to highly parallel scientific codes, where lower-precision storage both shrinks message sizes and further reduces communication overhead. While finer-grained tuning exists \cite{precimonious,tool-integration-mxp-source}, function-level granularity offers a practical balance by significantly reducing the search space.




Throughout, we eliminate regions unlikely to benefit from reduced precision. We adopt a staged strategy: converting the most time-critical inner routines to single precision first, then extending to outer bottlenecks where each stage includes the previous one. Finally, we implement the mixed-precision code and monitor accuracy, time-to-solution, and energy-to-solution on the LUMI system using SLURM’s energy accounting plugin and HPE Cray PM Counters \cite{crayx30, crayx30powermonitorandmanage}, guided by our CEEC Best Practice Guide \cite{ggedik_2024_13306639}.

%% file: reactor.tex
\input{side-by-side-tables}
\section{Reactor Simulator Benchmark}
\label{sec:reactor_simulator}
The Reactor Simulator~\cite{reactorsimulator,kahaner1989numerical} 
implements a probabilistic Monte Carlo application modeling interactions between neutron-source particles and a slab.
It emits particles iteratively, updates their states, and accumulates $total$-$energy$ in double precision while tracking outcomes with integer counters, making it a robust benchmark for mixed-precision techniques. 

Applying higher precision to whole application reduces forward error in $total$-$energy$ nearly linearly as mantissa bits vary from $3$ to $52$, regardless of the number of particles. This linear relationship is not trivial as highlighted in \cite{chen2024}.

Profiling reveal that shared math library calls (e.g., \texttt{sin\_\allowbreak{}fma()}, \texttt{exp\_fma()}) dominate execution. To evaluate precision impacts, we instrumented these functions using VPREC backend. We observed that the error plateaus after 17 bits, indicating FP32 accuracy is sufficient as shown in Fig. 1.

We proceed to instrument each function individually at single precision with VPREC backend, while leaving the others in double precision. By following this approach we explore the extent to which we can reduce precision. We notice that only 2 functions out of 10 have the potential to contribute to forward error under this approach. 

These insights guided our implementation of a five-stage mixed-precision workflow. In Stage 5, most functions run in single precision, with only accuracy-critical routines in double precision. Table~\ref{tab:gains_reactor_lumi_combined} summarizes time and energy savings on LUMI machine.

%% file: side-by-side-tables.tex
\begin{table*}[t]
  \centering
  \begin{minipage}[t]{0.48\textwidth}
    \centering
    \setlength{\tabcolsep}{4pt}
    \renewcommand{\arraystretch}{1.2}
    \begin{tabular}{| c | c | c | c | c | c | c |}
      \hline
      \textbf{Type} & \textbf{Stage 1} & \textbf{Stage 2} & \textbf{Stage 3} & \textbf{Stage 4} & \textbf{Stage 5} & \textbf{Double} \\
      \hline\hline
      Energy Median (J) & 1060 & 1220 & 1200 & 1250 & 1230 & 1700 \\
      \hline
      Energy Savings (\%)    & \bf{37.6\%} & 28.2\% & 29.4\% & 26.4\% & 27.6\% & - \\
      \hline\hline
      Time Median (s)    & 3.89 & 4.15 & 4.12 & 4.22 & 4.12 & 5.63 \\
      \hline
      Time Savings  (\%)   & \bf{30.7\%} & 26.2\% & 26.6\% & 25\% & 26.7\% & - \\
      \hline
      Error & 0 & 0 & 0 & $10^{-8}$ & $10^{-7}$ & - \\
      \hline
    \end{tabular}
    \caption{Energy (in Joules) and time-to-solution (in seconds) with Reactor Simulator with 10 million elements.}
    \label{tab:gains_reactor_lumi_combined}
  \end{minipage}\hfill
  \begin{minipage}[t]{0.48\textwidth}
    \centering
    \setlength{\tabcolsep}{4pt}
    \renewcommand{\arraystretch}{1.2}
    \begin{tabular}{| c | c | c | c | c |}
      \hline
      \textbf{Type} & \textbf{Stage 1} & \textbf{Stage 2} & \textbf{Stage 3} & \textbf{Double} \\
      \hline\hline
      Energy Median (J) & 1060 & 803  & 965  & 1080 \\
      \hline
      Energy Savings (\%)    & 1.8\% & \bf{25.6\%} & 10.6\% & - \\
      \hline\hline
      Time Median (s) & 3.53 & 2.71 & 3.20 & 3.96 \\
      \hline
      Time Savings (\%)    & 10.8\% & \bf{31.5\%} & 19.0\% & - \\
      \hline
      Error & $10^{-9}$ & $10^{-9}$ & $10^{-9}$ & - \\
      \hline
    \end{tabular}
    \caption{Energy (in Joules) and time-to-solution (in seconds) with LULESH with $20^{3}$ elements.}
    \label{tab:gains_lulesh_lumi_combined}
  \end{minipage}
  \vspace{-5mm}
\end{table*}

%% file: lulesh.tex
\vspace{-2.7mm}
\section{LULESH}
\enlargethispage{2.7mm}
\label{sec:lulesh_chapter}
LULESH (Livermore Unstructured Lagrangian Explicit Shock Hydrodynamics) is a proxy application developed by LLNL~\cite{LULESH:versions, LULESH2:changes, hydrochallenge} that replicates the computational behavior of a hydrodynamics code \cite{LLNL_LULESH_46c2a1d}. 
The simulation represents real-world HPC applications with its computational and memory access patterns.

With a mesh of $20^3$ elements, our experiments reveal that error decreases nearly linearly with increased precision. Notably, at 23 bits of mantissa (corresponding to FP32), we observe a stagnation at the timestep selection with Verificarlo, indicating a precision limit. We also observe that 10 functions out of 40 do not contribute  to end error of the $energy$ variable, which is the objective of the simulation to compute, when emulated in single precision separately while the rest of the application is kept in double precision.

Profiling revealed that a small number of routines accounted for a significant portion of the runtime. Functions requiring high numerical accuracy, such as \texttt{Time\allowbreak{}Increment()} and \texttt{Calc\allowbreak{}Position\allowbreak{}For\allowbreak{}Nodes()}, were retained in double precision. In contrast, performance-critical computations were selected for reduced precision to improve efficiency.

Mixed-precision optimization was carried out in three stages. In the first stage, only the core routine \texttt{Calc\allowbreak{}Elem\allowbreak{}FB\allowbreak{}Hour\allowbreak{}glass\allowbreak{}Force()} was converted to single precision; however, the element-wise casts and copies it imposed on its caller introduced overhead that yielded negligible speedup as seen in Table~\ref{tab:gains_lulesh_lumi_combined}. In the second stage, we extended this conversion to \texttt{Calc\allowbreak{}FB\allowbreak{}Hour\allowbreak{}glass\allowbreak{}Force\allowbreak{}For\allowbreak{}Elems()} by modifying its signature to accept single-precision inputs and relocating all cast/copy operations into its caller, \texttt{Calc\allowbreak{}Hour\allowbreak{}glass\allowbreak{}Control\allowbreak{}For\allowbreak{}El\allowbreak{}ems()}. This eliminated indirect-copy overhead and produced the largest performance gain. In the final stage, we implemented higher-level routines \texttt{Calc\allowbreak{}Hour\allowbreak{}glass\allowbreak{}Control\allowbreak{}For\allowbreak{}El\allowbreak{}ems()}, \texttt{Collect\allowbreak{}Dom\allowbreak{}ain\allowbreak{}Nodes\allowbreak{}to\allowbreak{}Elem\allowbreak{}Nodes()}, which showed zero error when instrumented alone in single precision with the help of VPREC backend, and \texttt{Volu\allowbreak{}Der()}  which significantly impacts runtime, its contribution to error is comparatively limited, thereby completing the full optimization.

%% file: conclusion.tex
\input{reactor-linear-relationship-picture}
\vspace{-1mm}
\section{Conclusion}
\enlargethispage{1mm}
  In this work, we applied a well-established methodology to two explicit solver use cases and showed that mixed-precision tuning is inherently application-specific. Our results demonstrate that selectively reducing precision in key computational kernels can significantly improve performance up to \% 30 and \%37 energy savings, while preserving acceptable numerical accuracy for both the Reactor Simulator and LULESH benchmarks.
  These results underscore the potential of mixed-precision as an effective approach to optimize scientific simulations for both performance and energy efficiency. Although challenges remain in narrowing the mixed‑precision search space and automating its implementation, upcoming work will address these aspects and evaluate our approach across a wider range of scientific applications. 




%% file: reactor-linear-relationship-picture.tex
\begin{figure}
    \centering
       \includegraphics[scale=0.35]{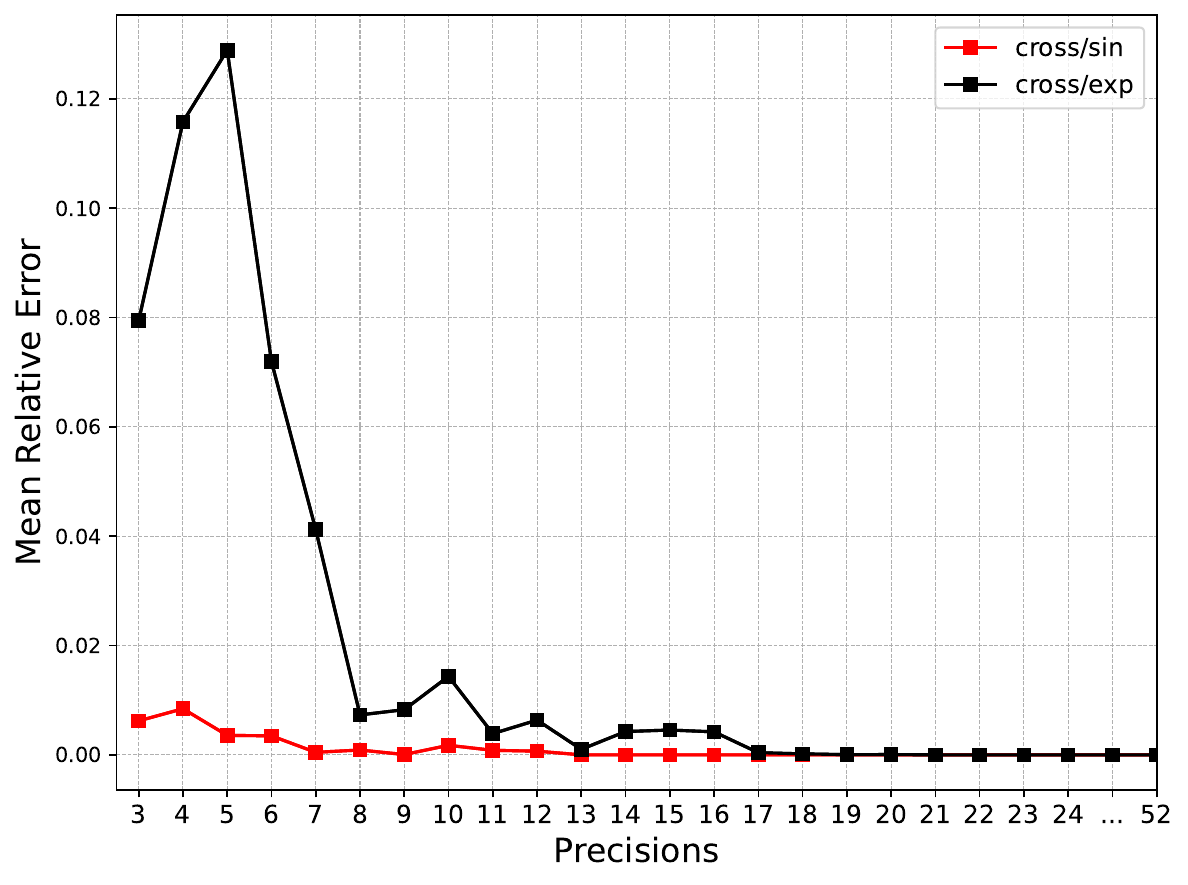}
    \caption{Mean absolute relative forward error for varying mantissa bits 
    $[3, 52]$ in \texttt{sin()} and  \texttt{exp()} in \texttt{cross()}.}
    \label{fig:sin_reactor_results}
    \vspace*{-5mm}
\end{figure}

%% file: measurement-meth-details.tex

All experiments were performed on the CPU partition of the LUMI system, where we compiled each benchmark with \texttt{g++} version 12.2.0 and the \texttt{-O2} optimization flag. To obtain statistically significant measurements, every application was run 32 times under the performance‐scaling governor, and both runtime (in seconds) and energy consumption (in joules) were recorded. Energy data were gathered via SLURM’s energy‐accounting plugin, which reads HPE Cray PM Counters through the Baseboard Management Controller (BMC) as detailed in our CEEC Best Practice Guide; nodes were dedicated exclusively to each job to eliminate interference from co‐scheduled workloads. CPU time was measured with the GNU \texttt{time} command, capturing only user and system time to reflect the precise CPU resources devoted to each process.

%% file: error-analysis-details.tex


To verify the numerical accuracy of our results, we recompiled the code using Verificarlo (version 1.0.0) with the \texttt{-O2} optimization flag to avoid unintended alterations from compiler optimizations. During standard runs, Verificarlo’s VPREC backend was employed with its default settings, preserving IEEE-754 binary64 precision and range. For mixed-precision experiments, we invoked VPREC with the \texttt{-precision-binary64=<desired\_mantissa\_bits>} option to emulate double precision at the specified mantissa width. This approach ensures that only the targeted precision changes, rather than other compiler transformations, impact our accuracy measurements. We quantify the error introduced by reduced precision using the mean absolute relative error.